\documentclass[twocolumn,pre,amsmath,amssymb,floatfix,superscriptaddress,showpacs]{revtex4}
\usepackage{graphicx,eucal,amsmath}
\pdfoutput=1
\begin{document}

\newcommand{\remove}[1]{}
\newcommand{\ie}{\emph{i.e.}}
\newcommand{\jmax}{j_{\rm max}}
\newcommand{\jfront}{j_{\rm front}}
\newcommand{\olc}{\overline{c}}

\author{Raissa M. D'Souza} 
\affiliation{Department of Mechanical and Aeronautical Engineering, University of California, Davis, CA 95616}
\affiliation{and the Santa Fe Institute, Santa Fe, NM 87501}
\author{Paul L. Krapivsky}
\affiliation{Department of Physics, Boston University, Boston, MA 02215}
\author{Cristopher Moore}
\affiliation{Computer Science Department, University of New Mexico, Albuquerque, NM 87131}
\affiliation{and the Santa Fe Institute, Santa Fe, NM 87501}

\title{The power of choice in network growth}
\begin{abstract}
The ``power of choice'' has been shown to radically alter the behavior of a number of 
randomized algorithms.  Here we explore the effects of choice on models of tree and network growth.  
In our models each new node has $k$ randomly chosen contacts, where $k > 1$ is a constant.  
It then attaches to whichever one of these contacts is most desirable in some sense, such as its 
distance from the root or its degree.  Even when the new node has just two choices, 
\ie, when $k=2$, the resulting network can be very different from a random graph or tree.  For instance, 
if the new node attaches to the contact which is closest to the root of the tree, the 
distribution of depths changes from Poisson to a traveling wave solution.  
If the new node attaches to the contact with the smallest degree, the degree distribution 
is closer to uniform than in a random graph, so that with high probability there are no nodes in the 
network with degree greater than $O(\log \log N)$.   Finally, if the new node attaches to the contact 
with the largest degree, we find that the degree distribution is a power law with exponent $-1$ 
up to degrees roughly equal to $k$, with an exponential cutoff beyond that; thus, in this case, 
we need $k \gg 1$ to see a power law over a wide range of degrees.
\remove{
a new node arriving and connecting to an existing node chosen at random, we consider the scenario  where $k>1$ existing nodes are chosen at random 
and then one of those selected as the connection node based on a simple network criteria.  We are particularly interested  in the ``power of two choices''---whether the resulting network properties vary dramatically if $k=2$ rather than $k=1$.  If the criteria of interest is minimizing network depth,  the depth distribution is Poisson distributed for $k=1$, while for $k=2$ it instead obeys a traveling wave solution.   With just a small amount of additional overhead, the depth distribution changes dramatically.  On the other hand, if the criteria is degree, no such change results. The degree distribution decays exponentially for all small values of $k$. Only once we allow $k\gg1$ does a power law regime appear (for degree $i <k$), meaning that generating a power law requires access to much information on the internal state of the system. 
}
  
  
\end{abstract}

\maketitle

\section{Formulation of the model}
\label{sec:intro}

Over the past decade, the ``power of choice'' has emerged as a theme in research on optimization and randomized algorithms~\cite{AzarBroder94, broder, adler, michael}.  Consider a random decision process. Typically at each step of the process a decision is reached by choosing one outcome at random and accepting this choice.  Now, rather then one random alternative being presented at each decision point, let a small set of randomly generated alternatives be presented, and let the best one be selected.  It has been shown that with as few as two alternatives at each decision point, the resulting properties of the process can be radically altered.  This was first explored in the context of load-balancing the allocation of jobs arriving at random times to a batch of processors.  With as few as two choices, the maximum load on any one processor drops dramatically from $O(\log N)$ to $O(\log \log N)$.   Increasing the number of choices beyond two only improves this by a constant factor, illustrating the ``power of two choices.''  

Here we explore the effect of choice on random network growth. 
Perhaps the simplest way to build a growing random network is to attach each new node  to an existing node which is chosen uniformly at random. This process 
generates {\em random recursive trees} which have been studied in great detail 
(see e.g.~\cite{mh,drmota,leader,sj} and references therein). Here we discuss a simple generalization: for each new node we choose $k>1$ existing `contact' nodes uniformly at random, select the `best' one according to some definition, and connect the new node to it.   This creates a random tree~\cite{note} whose statistics may be very different from those of a random recursive tree. 

We have to define, of course, the `quality' of the node so that we can choose the best one.  One natural definition of quality in a tree is distance to the root --- the closer to the root, the better, so that the new node attaches to whichever one of its contacts is closest to the root (and, if more than one contact has this smallest distance, we choose one of them randomly).  
This could correspond, for instance, to someone joining a hierarchical organization, and choosing to become a daughter node of whichever one of their $k$ contacts is highest up in the hierarchy. 

Another natural definition is to measure quality by degree of the contact node: for instance, to attach the new node to the contact node with highest degree, again breaking ties randomly.  Note that this is very different from the preferential attachment process~\cite{BA}, where the contact is selected from the entire graph with probability proportional to its degree.  This latter process requires complete knowledge of 
the degree of all existing nodes.  In contrast, our model assumes that the new node possesses only a small amount of local information, namely, the degrees of a small number of potential contacts.  This  brings us to another motivation for this work: the desire to understand the effects of limited, local information on network growth.  

For the smallest-depth model, we find a marked difference in behavior for $k \ge 2$ versus $k=1$.  The measure of interest in this case is the depth distribution (the fraction of nodes at each depth $j$).  For $k=1$, \ie, a random recursive tree, this distribution is Poisson.  For $k \ge 2$, the same Poisson distribution is observed for distances close to the root, however for larger distances the depth distribution obeys a traveling wave solution.  We also consider using {\em maximal} depth, rather than minimal depth,  as the contact node selection criterion and find a similar traveling-wave solution.

For the highest-degree model, we find that the degree distribution decays exponentially for degree $i>k$.  For $i<k$ the degree distribution exhibits power-law like behavior, thus in order to observe a power law for any substantial regime requires $k\gg 1$.  In other words, a large amount of (overhead/state/knowledge of the system) is required to achieve a power law distribution.  

Finally, in analogy to the above-referenced works on load balancing, the lowest-degree model achieves a degree distribution which is very close to uniform, in which the maximum degree in the entire graph is $O(\log \log N)$ as opposed to the maximum degree in a Poisson distribution, which is roughly $O(\log N)$.

\section{Smallest Depth} 

Let $N$ be the total number of nodes and $D_j(N)$ be the number of nodes at distance $j$ from the root.  By definition, $D_0(N)\equiv 1$, since the root is distance 0 from
itself. Thus $D_0(N)$ is a deterministic quantity,
while $D_j(N)$ with $1\leq j<N$ are random variables. We shall focus on their averages 
$Q_j(N)\equiv \langle D_j(N)\rangle$.  An average value provides a good description of a random variable when it is large and hence fluctuations are relatively small; we will see that this is 
indeed correct for $D_1(N)$. 

To set the stage we begin in Sect.~\ref{RRT} with the simpler case of random recursive trees, 
 for which everything is already known (see e.g.~\cite{GNR}). We then consider the influence of $2$ or more choices in Sect.~\ref{choice}.

\subsection{Random recursive trees and depth}
\label{RRT}

The quantity $D_j$ grows each time a node at distance $j-1$ is selected as the contact node.  The average depth distribution thus satisfies the master equation 
\cite{GNC}
\begin{equation}
\label{Qj}
Q_j(N+1)=Q_j(N)+\frac{1}{N}\,Q_{j-1}(N)
\enspace . 
\end{equation}
This equation is {\em exact} and it applies even for $j\!=\!0$ if we set
\mbox{$Q_{-1}(N)\equiv 0$}.  Using the recursive nature of~\eqref{Qj}, we first solve for
$Q_1(N)$, then $Q_2(N)$, {\it etc}.  This gives
\begin{equation}
\label{Qj-sum1}
Q_j(N+1)=\sum_{1\leq m_1< \cdots <m_j\leq N}\frac{1}{m_1\times \cdots \times m_j}
\enspace . 
\end{equation}
Equivalently, we can recast the $j$-fold sums into simple sums, although
the results look less neat.  For example,
\begin{subequations}
\begin{align}
    &Q_1(N)=H_{N-1}
    \label{Q1-RRT}\\
    &Q_2(N)=\frac{1}{2}\left[(H_{N-1})^2-H_{N-1}^{(2)}\right]
   \label{Q2-RRT}
\end{align} 
\end{subequations}
where $H_N^{(p)}=\sum_{1\leq n\leq N}n^{-p}$ are harmonic numbers.  The asymptotic behaviors of $H_N\equiv H_N^{(1)}$, $H_N^{(2)}$, and other harmonic numbers are well-known~\cite{knuth}, and the resulting asymptotics of the depth distribution are
\begin{eqnarray*}
Q_1(N+1)&=&\ln N+\gamma+\frac{1}{2N}-\frac{1}{12 N^2}+ \cdots \\
Q_2(N+1)&=&\frac{1}{2}\,(\ln N)^2+\gamma\,\ln N
+\frac{1}{2}\left[\gamma^2-\frac{\pi^2}{6}\right]+ \cdots ,
\end{eqnarray*}
where $\gamma \approx 0.577$ is the Euler-Mascheroni constant. Analogous results hold for $Q_j(N)$ for larger $j$.

If we merely want to establish the leading asymptotic behavior, we can
replace the summation in~\eqref{Qj-sum1} by integration.  This leads to
the simple result
\begin{equation}
\label{Qj-sol-cont}
Q_j(N)\to \frac{(\ln N)^j}{j!}
\end{equation}
showing that in the limit $N \to \infty$, the depth distribution is Poisson with mean $\ln N$.  
Alternatively, we can derive~\eqref{Qj-sol-cont} within a continuum approach by
replacing finite differences by derivatives in the $N\to\infty$ limit of~\eqref{Qj}. 
This procedure recasts discrete master equations into differential equations
\begin{equation}
\label{Qj-diff}
\frac{dQ_j}{dN}=\frac{1}{N}\,Q_{j-1}
\end{equation}
Solving~\eqref{Qj-diff} one recovers~\eqref{Qj-sol-cont}.

The normalization requirement $\sum_{j\geq 0}D_j(N)=N$ implies the sum rule for the averages
\begin{equation}
\label{Qj-sum}
\sum_{j\geq 0} Q_j(N)=N
\end{equation}
The continuum approximation~\eqref{Qj-sol-cont} agrees with the sum rule~\eqref{Qj-sum1} implying
that it well approximates the depth distribution in the entire range. We therefore use it to
find the depth of the recursive random tree. The depth is defined as the maximal 
$\jmax$. The criterion $Q_{\jmax}=1$ leads to an estimate~\cite{GNR}
\begin{equation}
\label{j-max}
\jmax=e\ln N
\end{equation}
It is possible to derive this result within the exact (discrete) approach and to
determine the fluctuations of $\jmax$.  
However, for our purposes~\eqref{j-max} is sufficient. 

\subsection{The model with $k=2$ choices}
\label{choice}

Now suppose the new node has $k=2$ choices.  In this case, 
we have $D_j(N+1)=D_j(N)+1$ 
if the two contact nodes have minimum depth $j-1$, or equivalently, if both of them have 
depth at least $j-1$, but if they do not both have depth greater than $j-1$.  The probability of this is 
\begin{align}
& N^{-2} \left[ \left(\sum_{i \geq j-1} D_i \right)^2 - \left( \sum_{i \geq j} D_i \right)^2 \right] 
\nonumber \\
= \; & N^{-2} \left( D_{j-1}^2+2D_{j-1}\sum_{i\geq j}D_i \right)
\enspace .
\label{growth} 
\end{align}
This leads to the exact recurrence
\begin{equation}
Q_j(N+1)=Q_j(N)+N^{-2}\left\langle D_{j-1}^2+2D_{j-1}\sum_{i\geq j}D_i\right\rangle 
\enspace . 
\nonumber
\end{equation}
Unfortunately, this is not very helpful since the average of the product of random quantities differs from the product of their averages, viz.\  
$\langle D_iD_j\rangle\ne \langle D_i\rangle\langle D_j\rangle$. One can, of course,
write down an exact recurrence for $\langle D_iD_j\rangle$, but this involves
third order moments  $\langle D_iD_jD_k\rangle$, and so on.  Thus the hierarchical nature of the governing equations does not allow us to obtain complete and rigorous results as is possible for the 
case $k=1$.
 
The cases of $j=1,2$ are exceptional and one can determine $Q_1$ and $Q_2$ analytically. For $j=1$ the analysis is especially simple since $D_0=1, \sum_{i\geq 1}D_i=N-1$, and the growth rate~\eqref{growth} simplifies to $[1+2(N-1)]/N^2$. Therefore the average number of the neighbors of the root grows according to an exact and closed recurrence
\begin{equation}
\label{Q1}
Q_1(N+1)=Q_1(N)+\frac{2N-1}{N^2}
\end{equation}
Solving~\eqref{Q1} subject to $Q_1(1)=0$ yields 
\begin{equation}
\label{Q1-sol}
Q_1(N)=\sum_{n=1}^{N-1}\frac{2n-1}{n^2}=
2H_{N-1}-H_{N-1}^{(2)}
\end{equation}
Similarly for $j=2$ we use relation $\sum_{i\geq 2}D_i=N-1-D_1$ and obtain
\begin{equation}
\label{Q2}
Q_2(N+1)=Q_2(N)+2\frac{N-1}{N^2}\,Q_1(N)-
\frac{\langle D_1^2(N)\rangle}{N^2}
\end{equation}
To obtain a closed recurrence for $Q_2$ we need to determine
$\langle D_1^2(N)\rangle$, the average of the square of the number 
of neighbors of the root. Then~\eqref{growth} leads to
\begin{equation*}
D_1(N+1)=
\begin{cases}
D_1(N)+1   & {\rm prob}\quad N^{-2}(2N-1)\cr
D_1(N)       & {\rm prob}\quad 1-N^{-2}(2N-1)
\end{cases} 
\end{equation*}
Squaring this equation and averaging we obtain
\begin{eqnarray*}
\langle D_1^2(N+1)\rangle&=&\left(1-\frac{2N-1}{N^2}\right)\langle D_1^2(N)\rangle\\
&+&\frac{2N-1}{N^2}[\langle D_1^2(N)\rangle+2Q_1(N)+1]\\
&=&\langle D_1^2(N)\rangle+2\frac{2N-1}{N^2}\,Q_1(N)+\frac{2N-1}{N^2}
\end{eqnarray*}
Rather than directly solving this recurrence, we can use it together with~\eqref{Q1} 
to establish a simpler recurrence for the variance  
$V_1(N)=\langle D_1^2(N)\rangle-\langle D_1(N)\rangle^2$. We find
\begin{equation}
\label{VN}
V_1(N+1)=V_1(N)+\frac{2N-1}{N^2}-\left(\frac{2N-1}{N^2}\right)^2
\end{equation}
which is readily solved to give
\begin{equation*}
V_1(N+1)=2H_N-5H_N^{(2)}+4H_N^{(3)}-H_N^{(4)}
\enspace . 
\end{equation*}
Thus $\langle D_1^2\rangle\ne \langle D_1\rangle^2$, yet the variance
is asymptotically $2\ln N$ and therefore fluctuations of the random variable
$D_1$ are indeed small compared to its average which grows as $2\ln N$,
see~\eqref{Q1-sol}. 

We determined $\langle D_1^2(N)\rangle=V_1(N)+Q_1^2(N)$ and therefore 
$Q_2$ satisfies a closed solvable recurrence~\eqref{Q2}. The solution reads
\begin{eqnarray*}
Q_2(N)&=&\frac{1}{2}\,[Q_1(N)]^2
-\frac{1}{2}\sum_{n=1}^{N-1}\left(\frac{2n-1}{n^2}\right)^2\\
&-&\sum_{n=1}^{N-1}\frac{V_1(n)+[Q_1(n)]^2+Q_1(n)}{n^2}
\enspace . 
\end{eqnarray*}

For $j\geq 3$, the problem becomes genuinely hierarchical and intractable. 
If we are seeking only the leading behavior, however, we can proceed. 
When $N\gg 1$ and $j$ is sufficiently small, namely such that  $\sum_{i \le j} Q_i \ll N$, 
we can replace the sum $\sum_{i\geq j}D_i$ by $N$ and the growth rate~\eqref{growth} 
by $2D_{j-1}/N$. Thus we arrive at a set of differential equations
\begin{equation}
\label{Qj-simple}
\frac{dQ_j}{dN}=2\,\frac{Q_{j-1}}{N}
\enspace . 
\end{equation}
Solving these equations we obtain
\begin{equation}
\label{Qj-sol}
Q_j(N)=\frac{(2\ln N)^j}{j!}
\enspace . 
\end{equation} 
We check the validity of this approximation by substituting it back into our assumption 
$\sum_{i \le j} Q_i \ll N$ which we used in the derivation of~\eqref{Qj-simple}.  
This suggests that~\eqref{Qj-sol} holds when $j < v \ln N$ (\ie, for small distances from the root) 
where $v$ is the smallest positive root of 
\begin{equation}
\label{v}
v\,\ln\left(\frac{2e}{v}\right)=1
\enspace . 
\end{equation}
We can write $v$ in terms of Lambert's function $W(x)$, defined as the root of $W e^W = x$:
\begin{equation}
\label{v:w}
v = -1/W_{-1}(-1/2e)
\end{equation}
where $W_{-1}$ denotes the $-1$st branch of the Lambert function.  
Numerically, $v=0.373365...$ 

When $j\geq v\ln N$ we cannot use~\eqref{Qj-sol}.  However, as long as $Q_j$ is much larger than 1, let us assume that the fluctuations in $D_j$ are small.  In that case we can replace averages $\langle D_j D_k\rangle$ by $Q_jQ_k$, and in this regime we obtain
\begin{equation}
\label{Qj-eq}
\frac{dQ_j}{dN}=N^{-2}\left( Q_{j-1}^2+2Q_{j-1}\sum_{k\geq j}Q_k \right)
\enspace. 
\end{equation}
\remove{
Since $\sum_{k\geq j}Q_k=N-\sum_{k<j} Q_k$, we can re-write~\eqref{Qj-eq} as
\begin{equation}
\label{Qj-rec}
\frac{dQ_j}{dN}=2\frac{Q_{j-1}}{N}
-\frac{Q_{j-1}^2}{N2^2}-2\frac{Q_{j-1}}{N^2}\sum_{k=0}^{j-2}Q_k
\end{equation}
which of course reduces to~\eqref{Qj-simple} when $Q_j\ll N$.  The manifestly recurrent 
form of~\eqref{Qj-rec} makes it useful for numerical treatment. For analytical work, it is more convenient 
}
It is convenient to introduce the cumulative variable
\begin{equation}
\label{q-def}
q_j=\frac{1}{N}\sum_{i \geq j} Q_i 
\end{equation}
that is, the average fraction of nodes whose depth is at least $j$. Summing~\eqref{Qj-eq} over
all $i \geq j$ we arrive at a neat recurrence
\begin{equation}
\label{qn}
\frac{d}{dN}\,N q_j=q_{j-1}^2
\enspace. 
\end{equation}
The form of this equation suggests the introduction of a new `time' variable
\begin{equation}
\label{time}
t=\ln N
\enspace. 
\end{equation}
This transformation recasts~\eqref{qn} into
\begin{equation}
\label{qnt}
\frac{d q_j}{dt}=-q_j+q_{j-1}^2
\end{equation}
which should be solved subject to the step function initial condition:
$q_j(0)=1$ for $j \leq 0$ and $q_j(0)=0$ for $j>0$. 

Equation~\eqref{qnt} has appeared in various contexts (see e.g.~\cite{extreme}) and 
while it is unsolvable, an asymptotic behavior of its solution is understood. 
In the long time limit, the solution approaches a `traveling wave' form, 
\begin{equation}
\label{wave}
q_j(t)\to q(j-vt)
\enspace. 
\end{equation}
Plugging~\eqref{wave} into~\eqref{qnt} one finds that $q(x)$ satisfies
\begin{equation}
\label{qx}
v\frac{d q}{dx}=q(x)-q(x-1)^2
\enspace. 
\end{equation}
The boundary conditions are 
\begin{equation}
\label{bc}
q(-\infty)=1,\quad q(+\infty)=0
\enspace. 
\end{equation}
The boundary-value problem~\eqref{qx}--\eqref{bc} is still intractable analytically.  However, the velocity $v$ can be determined even without a complete solution for $q(x)$. The method relies on the analysis of the tail region $x\to -\infty$. One notices that~\eqref{qx} admits an exponential solution in this region, 
\begin{equation}
\label{q-tail}
1-q(x)\propto e^{\lambda x} 
\quad {\mbox as} \quad 
x \to -\infty
\enspace. 
\end{equation}
Plugging this into~\eqref{qx} shows that the velocity $v$ is related to $\lambda$ via the dispersion relation~\cite{extreme}
\begin{equation}
\label{VL}
v=\frac{1-2 e^{-\lambda}}{\lambda}
\end{equation}
The maximum of $v=v(\lambda)$ is given by~\eqref{v} and it occurs at the largest positive root 
$\lambda$ of the transcendental equation $2(1+\lambda)=e^\lambda$.  This is 
\begin{equation}
 \lambda = -1-W_{-1}(-1/2e) 
\end{equation}
or numerically, $\lambda=1.67835...$  Comparing with~\eqref{v:w}, 
we see that $\lambda$ and $v$ 
are related as follows,
\begin{equation}
\label{lambda-v}
 \lambda = -1 + 1/v 
 \enspace .
\end{equation}

Strictly speaking, one can only assert that velocity does not exceed the maximum of~\eqref{VL}.  However, the so-called \emph{selection principle} tells us that this extremal value is realized for any initial conditions which vanish sufficiently rapidly at infinity. The selection principle has been rigorously proven for a few nonlinear parabolic partial differential equations. Yet heuristic arguments and numerical evidence indicate that the its range of applicability is much broader. This is reviewed in~\cite{van} in the context of partial differential equations and in~\cite{MK} in the context of difference equations. 

\begin{figure}[h]
\begin{center}
\resizebox{3in}{!}{\includegraphics{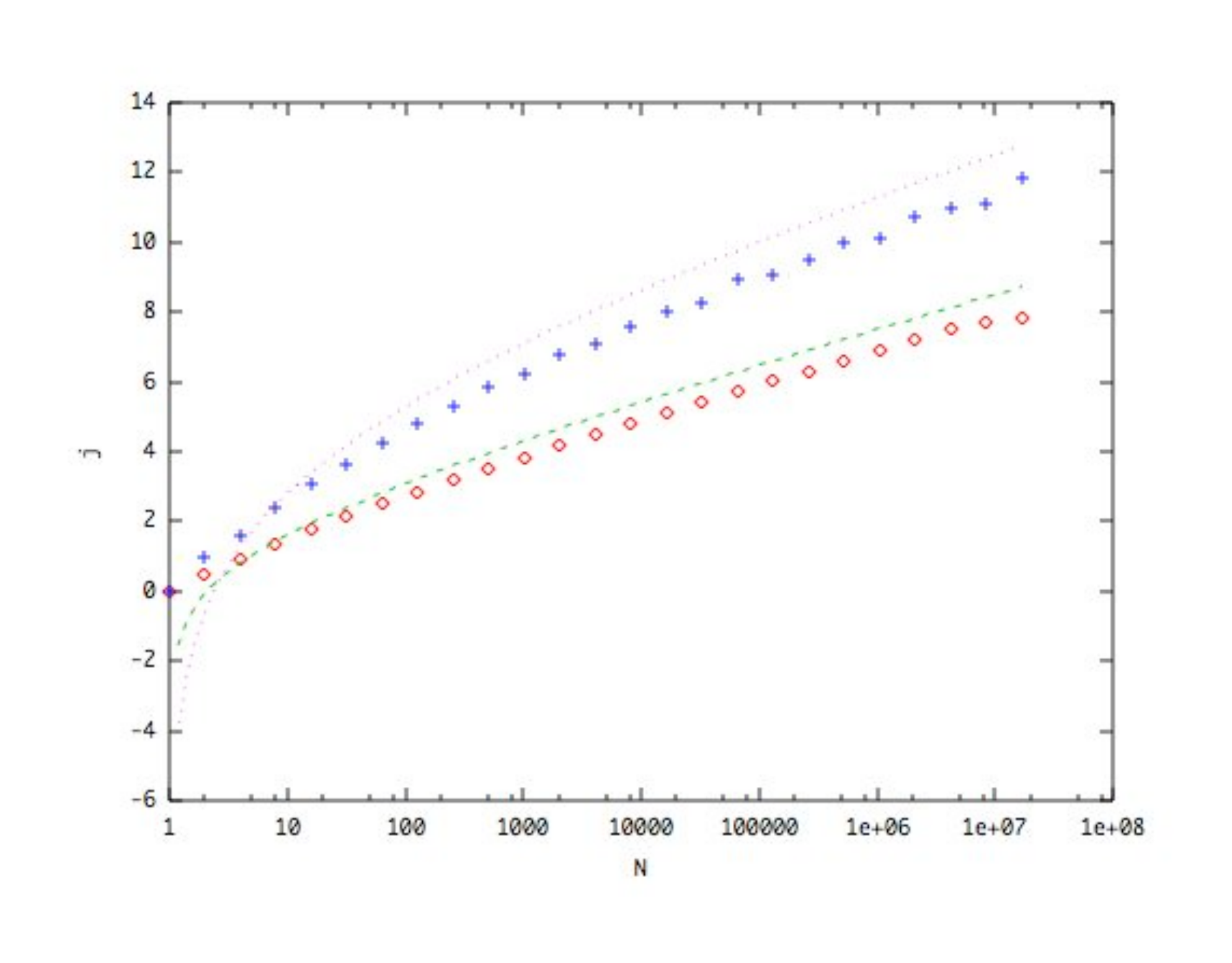}} 
\vspace{-3mm}
\caption{The average depth (circles) and maximum depth (crosses) of a tree with $k=2$, averaged over $10^3$ indepdent trials for each value of $N$, and (dashed) the expressions~\eqref{j-front-v} and~\eqref{j-max-v} for $\jfront$ and $\jmax$ respectively.}
\label{max-avg-k2}
\end{center}
\end{figure}

Thus there is a sharp front at depth $\jfront \approx vt = v\ln N$ to leading order, where the depth of most nodes in the tree is concentrated.  Furthermore, the width of this front remains finite even in the limit $
N\to\infty$.  It is also possible to compute the sub-leading correction to the position of the front~\cite{extreme}, giving an improved estimate of its location:
\begin{equation}
\label{j-front-v}
\jfront \approx v\ln N+\frac{3}{2\lambda}\,\ln\ln N
\enspace . 
\end{equation}
To estimate the \emph{maximum} depth $\jmax$, it is necessary to bound the tail of $q(x)$ in the positive direction $x \to +\infty$.  To do this, note that by definition $q(x)$ is monotonically decreasing, and by~\eqref{qx} this implies that 
\[ q(x) \le q(x-1)^2 \]
and therefore that this tail is doubly exponential,
\begin{equation}
 q(x) \propto e^{-A\cdot 2^x} 
\end{equation}
for some constant $A > 0$.  Setting $q(x) = 1/N$ then gives the estimate 
\begin{equation}
\label{j-max-v}
 \jmax \approx \jfront + \frac{\ln \ln N}{\ln 2} 
\end{equation}
minus a constant $C=\ln A / \ln 2$.  As shown in Fig.~\ref{max-avg-k2}, \eqref{j-front-v} and~\eqref{j-max-v} are indeed excellent estimates of the average and maximum depth respectively.

\subsection{The effect of choice}

At first sight, it seems that having two choices instead of one 
does not qualitatively affect the outcome, since the depth distributions~\eqref{Qj-sol-cont} and
\eqref{Qj-sol} both seem Poissonian, and both have typical depth $O(\log n)$. 
This is, however, an illusion.  First of all, the distribution~\eqref{Qj-sol-cont} for random 
recursive trees is indeed Poissonian while~\eqref{Qj-sol} is valid only for $j < v\ln N$.  
Secondly, while both types of trees have depth $O(\log n)$, choice causes the depth to be much more concentrated.  This is easiest to see if we consider the cumulative 
depth distribution~\eqref{q-def}.  For random recursive trees, $q_j(t)$ is asymptotically
\begin{equation}
\label{q-RRT}
q_j(t)=\frac{1}{2}\,{\rm erfc}\left(\frac{j-t}{\sqrt{2t}}\right)
\end{equation}
where ${\rm erfc}(z)$ is the error function 
\begin{equation}
\label{error}
{\rm erfc}(z)=\frac{2}{\sqrt{\pi}}\int_z^\infty d\eta\,e^{-\eta^2}
\end{equation}
Thus
\begin{equation*}
q_j(t)=
\begin{cases}
1  & j-t \ll -\sqrt{t} \cr
0  & j-t \gg +\sqrt{t}
\end{cases} 
\end{equation*}
The boundary layer where $q$ changes from one to zero is not a true front as its width grows with `time' as $\sqrt{t} \sim \sqrt{\ln N}$. 

On the other hand, for the model with choice the cumulative depth distribution has a traveling wave shape with a front of \emph{constant} width. Thus 
\begin{equation*}
q_j(t) =
\begin{cases}
1  & j-\jfront \ll -1\cr
0  & j-\jfront \gg +1
\end{cases} 
\enspace . 
\end{equation*}

\subsection{Multiple choices}

What if the new node has more than two choices?  The cases with $k\geq 3$ (with $k$ constant) 
are morally similar to the $k=2$ case: the cumulative depth distribution obeys the differential equation
\begin{equation}
\label{qnt-k}
\frac{d q_j}{dt}=-q_j+q_{j-1}^k
\enspace . 
\end{equation}
Transforming this to $q_j(t) = q(j-vt)$ as before, we obtain
\begin{equation}
\label{qx-k}
v\frac{d q}{dx}=q(x)-q(x-1)^k
\enspace. 
\end{equation}
The solution is again a traveling wave, whose velocity $v$ depends on $k$.  Assuming the selection principle, $v$ is the smallest positive root of
\begin{equation}
\label{v-m}
v\,\ln\left(\frac{ke}{v}\right)=1
\end{equation} 
which can be written in terms of Lambert's function as
\begin{equation}
\label{v-m:w}
v = -1/W_{-1}(-1/ke)
\enspace . 
\end{equation} 
Asymptotically, as $k$ grows we have
\begin{equation}
v \approx \frac{1}{\ln ke + \ln \ln ke} 
= \frac{1}{\ln k} \left( 1-O\!\left( \frac{\ln \ln k}{\ln k} \right) \right) 
\enspace .
\end{equation}
A more precise estimate for $\jfront$ is again given by~\eqref{j-front-v}, with $\lambda$ given by~\eqref{lambda-v}.  For $j \ll \jfront$, \eqref{Qj-sol-cont} and~\eqref{Qj-sol} generalize to
\begin{equation}
\label{Qj-sol-m}
Q_j(N)=\frac{(k\ln N)^j}{j!}
\enspace . 
\end{equation}
Finally, the tail of $q(x)$ is doubly exponential, 
\begin{equation}
 q(x) \approx e^{-A k^x} 
\end{equation}
and the maximum depth is given by
\begin{equation}
 \jmax \approx \jfront + \frac{\ln \ln N}{\ln k} 
 \enspace . 
\end{equation}

\section{Largest Depth}

We pause here to consider a model in which we reverse our definition of the `better' node, and attach each new node to the contact node which is furthest from the root.  If $k=2$, then we have $D_j(N+1)=D_j(N)+1$ whenever the maximum depth of the two nodes is $j-1$, and this occurs with probability
\begin{align}
& N^{-2} \left[ \left( \sum_{i=0}^{j-1} D_i \right)^2 - \left( \sum_{i=0}^{j-2} D_i \right)^2 \right] 
\nonumber \\
= \; & N^{-2}\left( D_{j-1}^2+2D_{j-1}\sum_{i=0}^{j-2} D_i \right) 
\enspace . 
\end{align}
For instance, the average number of the neighbors of the root grows according to
\begin{equation}
\label{Q1-max}
Q_1(N+1)=Q_1(N)+\frac{1}{N^2}
\end{equation}
and therefore
\begin{equation}
\label{Q1-max-sol}
Q_1(N) = H_{N-1}^{(2)}
\enspace . 
\end{equation}
Thus the average number of neighbors of the root does not diverge as in the smallest depth model, but instead approaches the constant $\zeta(2) = \pi^2/6$.  Generally, the behavior of $Q_j(N)$ for small $j$ is very different from~\eqref{Qj-sol}, viz.\ for $j=O(1)$ the average number of nodes of depth $j$ remains finite in the $N\to\infty$ limit. Therefore in contrast with the smallest depth model, the quantities $D_j(\infty)$ are not self-averaging when $j=O(1)$ and their averages do not characterize them.  Yet, the probability distribution 
\begin{equation}
\label{Ps-def}
P(s)={\rm Prob}[D_1(\infty)=s]
\end{equation}
can be determined. For instance, $D_1(2)=1$ and the probability that the root still has
one neighbor when the network size reaches $N$ is
\begin{equation*}
{\rm Prob}[D_1(N)=1]=\prod_{n=2}^{N-1}\left(1-\frac{1}{n^2}\right)
\end{equation*}
and therefore 
\begin{equation}
\label{P1}
P(1)=\prod_{n=2}^\infty\left(1-\frac{1}{n^2}\right)=\frac{1}{2}
\enspace . 
\end{equation}
Proceeding with this line of reasoning one obtains
\begin{equation*}
P(s+1)=\frac{1}{2}\sum_{2\leq n_1< \cdots <n_s}^\infty 
\frac{1}{(n_1^2-1)\times \cdots \times (n_s^2-1)}
\end{equation*}
which can be expressed as a sum involving the zeta function at positive integers. 

However, even though the $D_j$ are not self-averaging, there are many similarities between the smallest depth model and this one.  In particular, the cumulative depth distribution has a traveling wave shape~\eqref{wave}.  Indeed, after several mappings~\cite{MK} the model becomes identical to one which has appeared in studies of collision processes in gases~\cite{vvd}, fragmentation processes~\cite{km}, and other problems~\cite{extreme}.  If we define the cumulative variable as
\begin{equation}
q_j=\frac{1}{N}\sum_{i \geq j} Q_i 
\enspace , 
\end{equation}
then writing $q_j(t) = q(j-vt)$ gives~\eqref{qn}, \eqref{qnt} and \eqref{qx} again, but now with the boundary conditions 
\begin{equation}
q(-\infty) = 0 , \quad q(+\infty) = 1
\enspace .
\end{equation}
With these boundary conditions, \eqref{qx} admits a solution whose tail in the positive direction is exponential,
\begin{equation}
\label{mu-tail}
1-q(x) \propto e^{-\mu x} 
\quad {\mbox as} \quad 
x \to +\infty
\end{equation}
and the dispersion relation is now
\begin{equation}
\label{VM-largest}
v = \frac{2 e^\mu - 1}{\mu} \enspace . 
\end{equation}
The selection principle now suggests that $v$ is the \emph{minimum} of~\eqref{VM-largest}.  This is the larger of the two real roots of the transcendental equation~\eqref{v}, which is $v=4.31107...$  A more precise estimate of $\jfront$ is
\begin{equation}
\label{mu}
\jfront=v \ln N-\frac{3}{2\mu}\,\ln\ln N
\end{equation}
where $\mu=0.768039...$ is the larger root of $2(1-\mu)=e^{-\mu}$.  

More generally, for $k > 2$ the velocity $v$ is the larger real root of~\eqref{v-m}, or 
\begin{equation}
v = -1/W_{1}(-1/ke)
\end{equation} 
which, as $k$ grows, approaches 
\begin{equation}
v \approx ke - 1
\enspace . 
\end{equation}
The position of the front is given by~\eqref{mu} with
\begin{equation}
 \mu = 1-1/v \enspace . 
\end{equation}
Finally, since the tail of $q(x)$ in the positive direction is given by~\eqref{mu-tail}, setting $q_j = 1-1/N$ gives the following estimate of the maximum depth, 
\begin{equation}
\jmax = \jfront + \frac{1}{\mu} \ln N
\enspace . 
\end{equation}
Note that, unlike the minimum depth model, $\jmax-\jfront$ is $O(\log N)$ instead of $O(\log \log N)$, since the tail~\eqref{mu-tail} is exponential rather than doubly exponential.

\section{Highest Degree}\label{sec-highdegree}

We now consider a model in which quality is measured not by depth, but by the degree of the contact node --- the higher the degree, the better.  As we will show below, in this case the degree distribution exhibits a power law up to degree $j \sim k$, beyond which it decays exponentially.  Therefore, in this model we need a large number of choices, $k \gg 1$, in order to observe a power law over a wide range of degrees.

\subsection{Recurrence for the degree distribution}

We start by writing a master equation for the degree distribution of the network.  We add one node at each step, so at time $t$ there are $t$ nodes in the network.  Let $N_i(t)$ be the number of nodes which have degree $i$ at time $t$, and let $C_i(t) = \sum_{j=1}^i N_j(t)$ be the corresponding total number of nodes of degree $i$ or less at time $t$.  Normalizing these numbers, let $a_i(t)=N_i(t)/t$ be the fraction of nodes which have degree $i$, and let $c_i(t) = \sum_{j=1}^i a_j(t) = C_i(t)/t$ be the corresponding cumulative distribution.  

At each iteration, we choose $k$ contact nodes at random from the $t$ existing nodes, and connect the new node to the contact node of highest degree, with ties broken randomly.  The evolution of the expected cumulative degree distribution can can be written as, for all $i \ge 1$, 
\begin{equation}
C_i(t+1) = C_i(t) + 1 - \left( c_i(t)^k - c_{i-1}(t)^k \right)
\enspace ,
\label{discrete-recur}
\end{equation}
since $C_i$ increases by $1$ for each new node added, and decreases precisely when the new node connects to a node of degree $i$.  This latter event occurs when all $k$ nodes have degree $i$ or less, but not all have degree $i-1$ or less.  Writing $c_i(t) = C_i(t)/t$ and making the assumption that a steady-state limit exists, we obtain the recurrence
\begin{equation}
c_i = 1 - (c_i^k - c_{i-1}^k) \enspace .
\label{recurrence}
\end{equation}
We note that in the case $k=1$, where there is no choice, the solution to~\eqref{recurrence} is simply
\begin{equation}
c_i = 1 - 2^{-i} \;\mbox{ and }\; a_i = 2^{-i} 
\label{randtree}
\end{equation}
which is the degree distribution of a random recursive tree.

\subsection{The model with $k \ge 2$ choices} 

\begin{figure}[tb]
\begin{center}
\resizebox{2.75in}{!}{\includegraphics{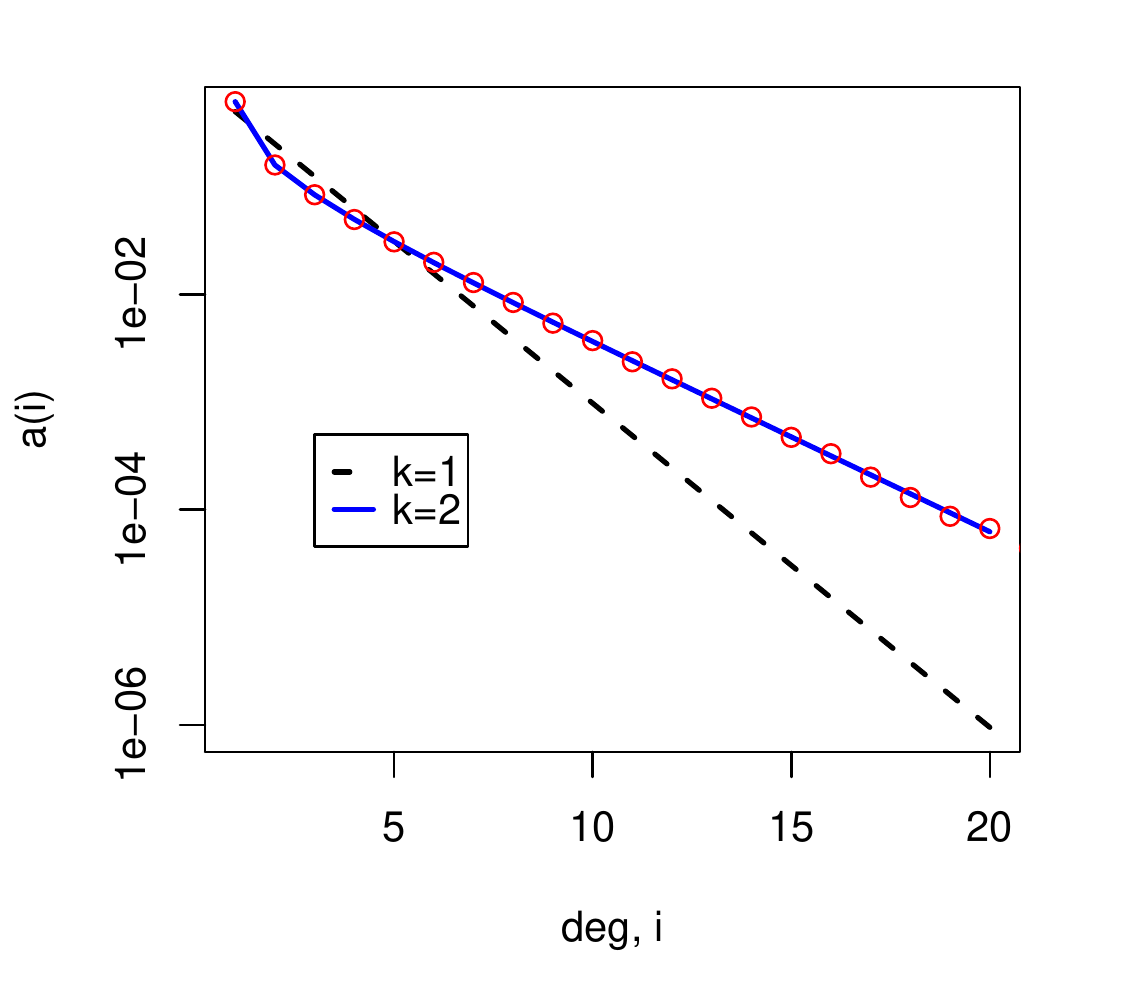}} 
\vspace{-3mm}
\caption{The degree distribution, $a_i$, for the highest degree model, for both $k=1$ and $k=2$. The points at data from numerical simulation of the model with $k=2$.}
\label{k1k2degree}
\end{center}
\end{figure}

We are particularly interested in the behavior for small $k$. Recall that the ``power of choice'' comes from situations where results vary dramatically if $k=2$ rather than $k=1$.  For $k\ge 2$ we can solve~\eqref{recurrence} analytically only in the regime $i\gg 1$ as discussed in detail below. Yet, for $k=2$, equation \eqref{recurrence} is very easy to solve numerically as it reduces to the quadratic equation:
\begin{equation}
c_i^2 + c_i - (1+c_{i-1}^2) = 0.
\end{equation}
Figure~\ref{k1k2degree} is a plot of the degree distribution, $a_i$, for $k=1$ and $k=2$.  
Recall $a_i = c_i - c_{i-1}$. 
The data points are from a numerical simulation with $k=2$, grown to size $1\times 10^6$ nodes. Note the excellent agreement.  Though the distribution for $k=2$ decays less slowly than $k=1$ both exhibit exponential decay, thus the nature of the solution is not altered with such minor amounts of choice.  

\begin{figure}[tb]
\begin{center}
\resizebox{2.75in}{!}{\includegraphics{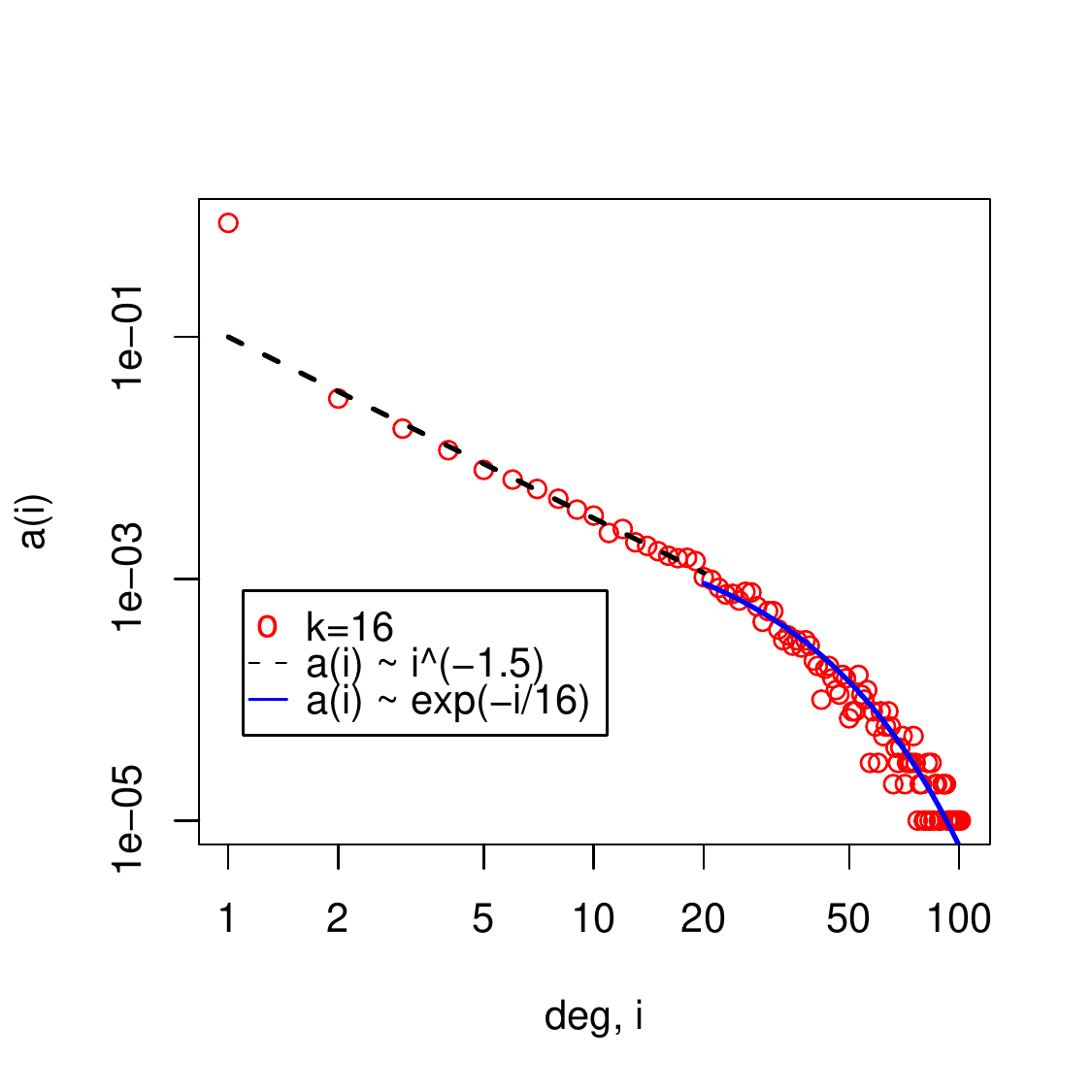}} 
\vspace{-3mm}
\caption{Numerical simulation results for $k=16$.  Note that for $i<k$ we observe $a_i \sim i^{-1.5}$, while for $i>k$ we observe $i \sim e^{-i/k}$.}
\label{k16degree}
\end{center}
\end{figure}

From numerical simulation with $k\ge 2$ we find different behaviors for $i > k$ than for $i < k$ (see Fig.~\ref{k16degree}).  For degree $i > k$ we observe $a_i \sim \exp(-i/k)$.   For $i < k$ we observe what appears to be a power law in that regime, $a_i \sim k^{-\gamma}$, with $\gamma \approx 1.5$.  The largest $k$ we simulated was $k=32$, hence the ``power law'' regime is quite small.  Rather than computer simulation, we can look at the asymptotic limits of~\eqref{recurrence} and arrive at these similar
results in the limit $i\gg 1$ and $k\gg 1$.  Note, the asymptotic limit will give $\gamma=1$, and we can attribute the difference with numerical results to finite size effects in simulation.  

\subsection{Asymptotic limits}

In the asymptotic regime $i \gg 1$ we write $c_i = 1 -\epsilon_i$ and assume that 
$\epsilon_i \ll 1$.  To first order, $c_i^k = 1- k\epsilon_i$.  Simplifying~\eqref{recurrence}, we find $(k+1)\epsilon_i=k\epsilon_{i-1}$ and therefore
\begin{equation}
\label{recur:asymp}
 1-c_i = A_k\left(\frac{k}{k+1}\right)^i \quad{\rm when} \quad i\gg 1,
\end{equation}
where $A_k$ is a constant depending on $k$.  We argue below that 
\begin{equation}
\label{A}
A_k\sim k^{-1}\quad{\rm as}\quad k\to\infty
\end{equation}

In the rest of this section we always assume that $k \gg 1$. Let us start with nodes of degree one (which are often called `leaves'). In this case we have $c_1=a_1$ and equation~\eqref{recurrence} reduces to
\begin{equation}
\label{eq:leaf}
 a_1 = 1-a_1^k.
\end{equation}
Writing 
\begin{equation}
\label{a1:seek}
a_1 = 1-\frac{W}{k}
\end{equation}
and assuming that $W \ll k$ yields $a_1^k = e^{-W}$.  This allows us to recast~\eqref{eq:leaf} into 
\begin{equation}
\label{Lambert}
W e^W=k
\end{equation}
so $W$ is Lambert's function $W(k)$.  For large $k$, we have $W(k) \approx \ln k$, justifying our assumption that $W \ll k$. 
Thus almost all nodes are leaves: the fraction of nodes whose degree exceeds one is $1-a_1 = W(k)/k \approx (\ln k)/k$. 

Analyzing~\eqref{recurrence} for $i=2,3,\ldots$ one finds that the following ansatz is useful:
\begin{equation}
\label{c:seek}
c_i=1-\frac{W-w_i}{k}
\end{equation}
Plugging~\eqref{c:seek} into~\eqref{recurrence} we obtain
\begin{equation}
\label{wwW}
1+e^{w_{i-1}}-e^{w_i}=W^{-1}w_i
\end{equation}
Since $W\to\infty$ as $k\to\infty$, Eq.~\eqref{wwW} simplifies to
\begin{equation}
\label{ww}
1+e^{w_{i-1}}-e^{w_i}=0
\end{equation}
whose solution (satisfying $w_1=0$) is $w_i=\ln i$. Plugging this to~\eqref{c:seek} we find that $a_i=c_i-c_{i-1}$ is given by 
\begin{equation}
\label{ai:small}
a_i=k^{-1}\ln\left(\frac{i}{i-1}\right)\quad{\rm when}\quad 2\leq i\ll k
\end{equation}
The upper bound $i\ll k$ is necessary since we can use~\eqref{ww} instead of
\eqref{wwW} only when $w_i\ll W$ which is equivalent to $\ln i\ll \ln k$. Note that we can further simplify~\eqref{ai:small} when $i\gg 1$, viz.
\begin{equation}
\label{ai:sol}
a_i=\frac{1}{k}\cdot\frac{1}{i}\quad{\rm when}\quad 1\ll i\ll k
\end{equation}
Thus up to a crossover at $i=k$, the degree distribution exhibits an algebraic behavior $a_i\sim i^{-1}$ with unusually small exponent. 

The derivation of~\eqref{recur:asymp} actually holds when $i\gg k$. Using 
\eqref{recur:asymp} we compute  $a_i=c_i-c_{i-1}$ to give
\begin{equation}
\label{ai:asymp}
 a_i = k^{-1}A_k\left(\frac{k}{k+1}\right)^i \quad{\rm when} \quad i\gg k
\end{equation}
The regions of the validity of~\eqref{ai:sol} and~\eqref{ai:asymp} do not formally overlap. 
It is reasonable to assume, however, that they remain qualitatively correct. Then 
from Eq.~\eqref{ai:sol} we obtain $a_k\sim k^{-2}$ while Eq.~\eqref{ai:asymp} leads to 
$a_k\sim k^{-1}A_k$. Matching this values we confirm the announced 
asymptotic of the amplitude,  Eq.~\eqref{A}.  Furthermore, we find 
\begin{equation}
a_i \sim \left(\frac{k}{k+1}\right)^i  \approx e^{-i/k} \quad{\rm when}\quad 1\ll k\ll i.
\end{equation}

\section{Lowest Degree}

There are situations where one wants to ensure that all nodes have low degree, for instance consider the case of load-balancing discussed in Sec.~\ref{sec:intro}.  Thus the final variant we consider is when an incoming node connects to the target node of {\em lowest} degree.  

\subsection{Recurrence for the degree distribution}

As in Sec.~\ref{sec-highdegree}, we begin by writing the master equation for the degree distribution of the network.  Again let $N_i(t)$ be the number of nodes which have degree $i$ at time $t$, and now let $\overline{C}_i(t) = \sum_{j\geq i} N_j(t)$ be the corresponding total number of nodes of degree $i$ or greater at time $t$.  Normalizing, let $a_i(t)=N_i/t$ and let $\overline{c}_i(t) = \sum_{j\geq i} a_j(t) = \overline{C}_i(t)/t$ be the complementary cumulative distribution. 

At each iteration, we choose $k$ contact nodes at random from the $t$ existing nodes, and connect the new node to the contact node of {\it lowest} degree, with ties broken randomly.  The evolution of the expected complementary cumulative degree distribution can can be written, for all $i > 1$, as  
\begin{equation}
\overline{C}_i(t+1) = \overline{C}_i(t) +  \left[\overline{c}_{i-1}(t)^k - \overline{c}_{i}(t)^k\right]
\enspace ,
\label{min-discrete-recur}
\end{equation}
since $\overline{C}_i$ increases precisely when the new node connects to a node of degree $i-1$.  This event occurs when all $k$ nodes have degree $i-1$ or greater, but not all have degree $i$ or greater.  Writing $\overline{c}_i(t) = \overline{C}_i(t)/t$ and making the assumption that a steady-state limit exists, we obtain the recurrence
\begin{equation}
\overline{c}_i = \overline{c}_{i-1}^k - \overline{c}_{i}^k \enspace .
\label{min-recurrence}
\end{equation}
We note that in the case $k=1$, where there is no choice, the solution to~\eqref{min-recurrence} is simply
\begin{equation}
\overline{c}_i = 2^{-{(i-1)}} \;\mbox{ and }\; a_i = 2^{-i} 
\label{compl-randtree}
\end{equation}
which, as~\eqref{randtree}, is the degree distribution of a random recursive tree.

\subsection{The model with $k\ge 2$ choices}
\begin{figure}[htb]
\begin{center}
\resizebox{2.75in}{!}{\includegraphics{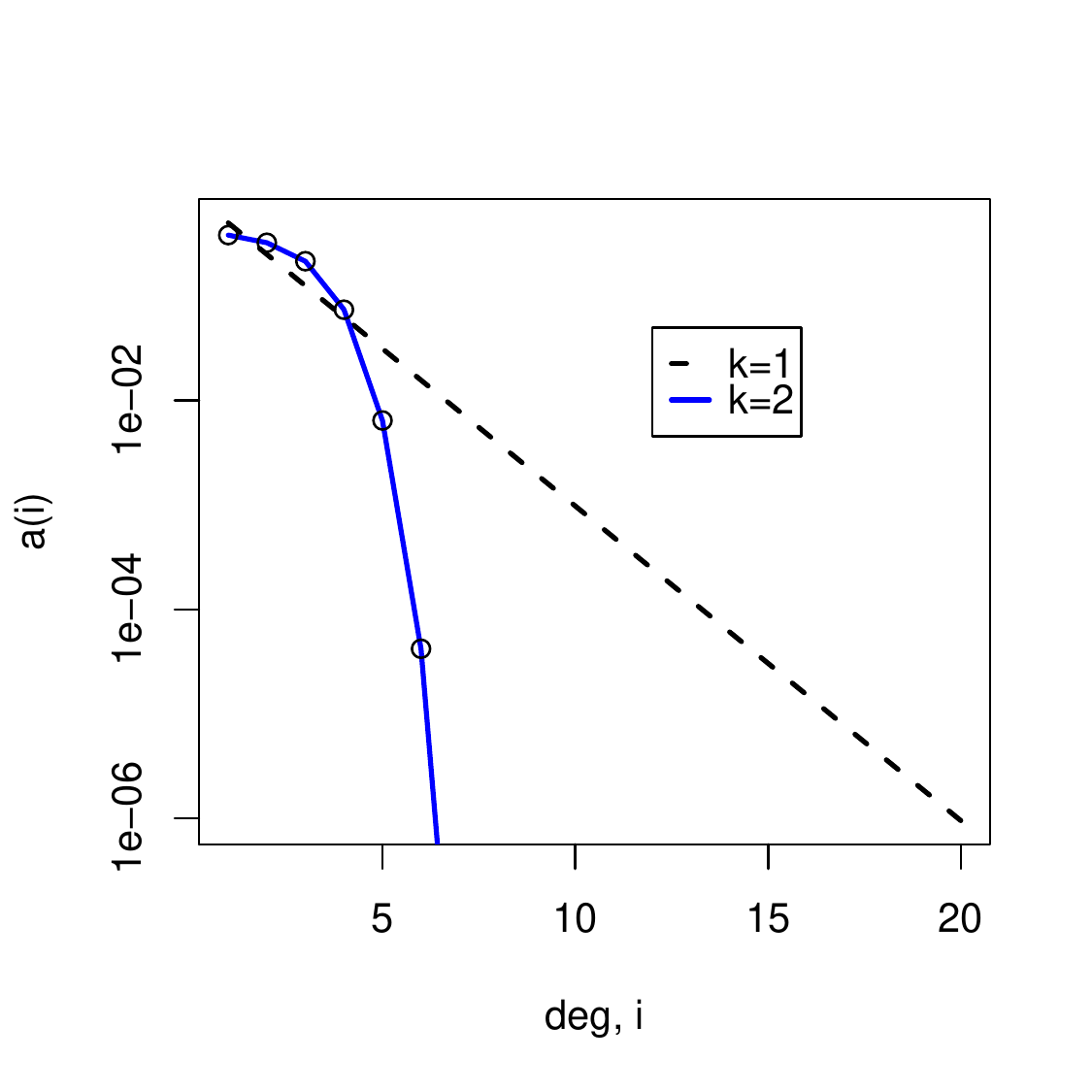}} 
\vspace{-3mm}
\caption{The degree distribution, $a_i$, for the lowest degree model, for both $k=1$ and $k=2$. The points at data from numerical simulation of the model with $k=2$.}
\label{k1k2min}
\end{center}
\end{figure}

For $k=2$, \eqref{min-recurrence} is very easy to solve numerically as it reduces to the quadratic equation:
\begin{equation}
\overline{c}_i^2 + \overline{c}_i - \overline{c}_{i-1}^2 = 0 \enspace .
\end{equation}
Figure~\ref{k1k2min} is a plot of the degree distribution, $a_i$, for $k=1$ and $k=2$.  
Recall here, $a_i = \overline{c}_i - \overline{c}_{i+1}$. 
The data points are from a numerical simulation with $k=2$, grown to size $1\times 10^6$ nodes. Note the excellent agreement.  With minor choice, the degree distribution is radically altered.  

For all $k\ge 2$ we can show the upper bound on the maximum degree is $O(\log \log N)$ using a method similar to that in \cite{mitzenmach01}.  
From \eqref{min-recurrence}, for $i \ge 3$ we obtain the upper bound, $\olc_i \le \olc_{i-1}^k$, and by recursion:
\begin{equation}
\olc_i \le \olc_{i-1}^k \le \olc_{2}^K,
\end{equation}
where $K = k^{(i-2)}$. Since $\olc_2 < 1$, $\olc_i$ decreases doubly-exponentially.  To find $i_{\rm max}$, the typical largest degree present after addition of $N$ nodes, we set $\olc_i = 1/N$. Solving this relation we find: 
\begin{equation}
i_{\rm max} \le \log_k \log_{1/\olc_2} N = O(\log \log N).
\end{equation}

\section{Discussion}

We explore the ``power of choice" in network growth by introducing a minimalist generalization of random recursive trees.  At each decision point $k>1$ choices are presented and the most desirable one selected.  If the criteria is to minimize or maximize network depth, a small amount of choice has a dramatic effect.   For $k=1$ the depth distribution decays with a Poisson behavior.  For $k\ge 2$ this Poisson decay is seen for distances close to the root, but for further distances, the depth distribution obeys a traveling wave behavior. If the criteria instead involves node degree, we must distinguish the maximum degree model from the minimum degree one.  For minimum degree, choice has a dramatic effect.  Going from $k=1$ to $k=2$ the degree distribution changes from geometric decay to double-exponential decay (and hence the maximum degree observed in the network changes from $O(\log N)$ to $O(\log \log N)$).  In contrast, for maximum degree, a large number of choices, $k \gg 1$, must be allowed before a change from the $k=1$ behavior is observed. The degree distribution decays exponentially for all small values of $k$.  Once $k\gg1$ a power law distribution results for nodes of degree $i < k$, while for nodes of degree $i > k$ the distribution decays exponentially.  

We established many results about the depth distribution. Some of them are exact,
others (namely the assumption that the maximum allowed value of velocity is realized, employed at the end of Sec.~\ref{choice}) utilize a selection principle which is not rigorously established for~\eqref{qnt}. There is no doubt of the validity of this principle in a broad range of contexts, and there is firm numerical support of all analytical results derived herein. 




\medskip
{\bf Acknowledgments.}
P.L.K. is thankful to CNLS (Los Alamos National Laboratory) for hospitality during the initial stage of this research, and to Renaud Lambiotte for interesting correspondence. C.M. is supported by NSF grant CCF-0524613 and ARO  
contract W911NF-04-R-0009


\begin{thebibliography}{99}
\bibitem{note}
     We always assume that we start with a tree; we further assume (if not stated
     otherwise) that the initial network is a single node which is the root.
\end{thebibliography}
\end{document}